# Accelerated structure-stability energy-free calculator

*Alexandre Boucher,[a] Cameron Beevers,[a] Bertrand Gauthier,[b] and Alberto Roldan[a]\**

[a] *Cardiff Catalysis Institute, School of Chemistry, University of Cardiff, Main Building, Park Pl, Cardiff CF10 3AT, Cardiff*

[b] *School of Mathematics, Cardiff University, Abacws building, Senghennydd Rd, Cardiff, CF24 4AG*

**Abstract**

Computational modeling is an integral part of catalysis research. With it, new methodologies are being developed and implemented to improve the accuracy of simulations while reducing the computational cost. In particular, specific machine-learning techniques have been applied to build interatomic potential from *ab-initio* results. Here, We report an energy-free machine-learning calculator that combines three individually trained neural networks to predict the energy and atomic forces of particulate matter. Three structures were investigated: a monometallic nanoparticle, a bimetallic nanoalloy, and a supported metal crystallites. Atomic energies were predicted via a graph neural network, leading to a mean absolute error (MAE) within 0.004 eV from Density Functional Theory (DFT) calculations. The task of predicting atomic forces was split over two feedforward networks, one predicting the force's norm and another its direction. The force prediction resulted in a MAE within 0.080 eV/Å against DFT results. The interpretability of the graph neural network predictions was demonstrated by underlying the physics of the monometallic particle in the form of cohesion energy.

1. Introduction

Catalysis is crucial to today's economy and achieving a sustainable future. Since its discovery in the 1800s,[1] heterogeneous catalysis has played a significant role in the chemical industry and applications for clean energy[2-4] and environmental control.[5] Supported metal nanoparticles (NPs) are frequently used as a type of heterogeneous catalyst that reduces the amount of metal required while generally improving catalytic activity.[6-8] Many experimental and computational efforts have been dedicated to revealing the morphologies of these nanoparticles and their interaction with the support, as these are important factors to consider in rationalizing the NP's chemical activity.[9-24]

The macroscopically observed properties of a catalyst derive from the collection of individual particles of various shapes and sizes under given system conditions; the collective of particles is commonly called the metastable ensemble.[25-28] To predict trends from a group of particles forming a heterogeneous catalyst, computational scientists need to sample their chemical space to account for relevant structures with enough stability to persist under reaction conditions. The stability of these nanoscale structures is often evaluated using density functional theory (DFT) or interatomic potentials (IP).[29] This is problematic because small NPs of size on the nanometre order, known as clusters, have a large number of low-lying energy configurations,[30,31] making the number of calculations required to map their stabilities overly demanding.[32,33] The number of structures with energies close to the global minimum (the most stable) increases exponentially with particle size.[34] Furthermore, the energy contribution from each atom depends on its position in the supported cluster, i.e., the surrounding chemical environment. Therefore, the parameterization and formulation of standard IP make them fundamentally unable to describe accurately the energies of supported clusters, whose properties are not scalable to the bulk.[29,34]

Recent progress in predicting potential energy surfaces (PES) is based on representative datasets of the spanned chemical space interpreted through machine learning (ML) architectures leading to neural network interatomic potentials (NNIPs) with near-DFT accuracy.[35-42] ML has proven to be a powerful tool for accelerating computational research in association with already existing global optimization methods.[35,36,41,43-

[48] Once constructed, NNIPs allow for highly efficient computation at a cost up to 1000 times cheaper than an accurate DFT calculation.

In the present work, we introduce an innovative tool for predicting the energies and forces of individual atoms forming a gas-phase or supported cluster. The approach combines state-of-the-art graph neural networks in an energy-free approach, where distinct neural networks compute the atomic energies independently from the forces. The investigated systems are gas-phase palladium and AuPd bimetallic clusters and supported palladium clusters on α-silica. The predictions reached near-DFT accuracy at a small fraction of its computational cost. The resulting set of neural networks predicting atomic energies and forces was organized in a specific architecture called a machine-learning calculator (ML-calculator). The ML-calculator forms an autonomous ML-based tool that can be coupled with existing DFT algorithms to accelerate the calculation rate or work on its own as an independent NNIP for geometry optimization.

## 2. Method

### 2.1. Density functional theory calculations

All calculations performed to generate the required datasets, i.e., geometry optimizations of Pd-pure and AuPd-alloy clusters in gas-phase and Pd supported on α-$SiO_2$(001), were carried out using spin-polarized Density-Functional Theory (DFT) as implemented within the Vienna *Ab-initio* Simulation Package (VASP).[49,50,51] The revised PBE functional from Perdew, Burke, and Ernzerhof (RPBE)[52,53] was used to calculate the exchange-correlation energy. The Projected-Augmented Wave (PAW) pseudopotentials were employed to describe core electrons.[51,54] Dispersion corrections were included through Grimme's dispersion correction scheme, DFT-D3.[55] The plane-wave kinetic cut-off was set to 500 eV, the electronic energy convergence threshold set to $1 \times 10^{-7}$ eV, and the ionic convergence to 0.04 eV/Å. Gaussian smearing was employed to describe the distribution of electrons around the Fermi level, with a smearing parameter of 0.2 eV for pure metallic structures and 0.1 eV for $SiO_2$ surfaces and supported clusters.

Supported Pd NPs on the α-$SiO_2$(001) surface were modeled using a *p(2x2)* supercell preventing the interactions of supported metal atoms with periodic images. The silica support contained 3 $SiO_2$ layers and all dangling bonds were saturated with hydroxyl groups. Only the surface hydroxyl groups and the palladium cluster were relaxed during geometry optimization (further details in Supporting Information (SI)). A vacuum layer of at least 10.0 Å was placed perpendicularly to the surface. Calculations were performed using a k-points density of 0.2 points/Å.

## 2.2. Datasets

Creating a reference dataset representative of the targeted PES is crucial to training a machine-learning neural network (NN). Due to the cost of running DFT calculations, the size of the dataset should be kept as small as possible while ensuring the integrity of the chemical environments relevant to the PES. Three datasets were prepared and exploited in the present work. Pd-pure, AuPd-alloy, and Pd-silica datasets

*Pd-pure:* The first dataset covers gas-phase Pd structures containing up to 55 atoms. It contains a total of 439 distinct structures taken from the literature and built from chemical intuition.[21,56–59] It is paramount to include structures outside the potential energy minima to improve the NN's versatility and capability, particularly for tasks such as geometry optimization.

*AuPd-alloy:* The dataset contains 116 Pd-pure and 45 Au-pure gas-phase structures derived from the literature, with sizes ranging from 17 to 34 atoms.[60,61] Besides, the dataset contains 93 structures of bimetallic AuPd gas-phase with various ratios, containing 19 to 27 atoms, in random, Janus, and core-shell arrangements.

*Pd/SiO$_2$(001):* The dataset combines the gas-phase Pd-pure dataset (439 structures) with 66 structures of up to 8 Pd atoms supported on the α-silica slab. It also contains 7 bare silica structures derived from the pristine α-silica(001) surface: Bare fully hydroxylated surface, shrunk along the axis perpendicular to the slab, only top -OH groups shrunk and elongated and up to 3 -OH groups missing.

## 2.3. Capturing the atomic environment

The atomic environment of each atom in the dataset was converted into a processable vector. This process, called atomic fingerprinting, has been developed over the years in multiple flavors, such as symmetry functions or smooth overlap of atomic position (SOAP).[62–64] The fingerprint developed in the present work extracted local and non-local data from the atomic structure. Local information was obtained using the G2 and G3

**4 |** P a g e

symmetry functions, capturing radial and angular features, respectively, as introduced by Behler.[65,66] These functions have been extensively described in the literature and successfully employed to generate multiple neural network interatomic potentials (NNIPs).[36,38–40,64–66] Non-local information was expressed through the $G^2$ and $G^3$ functions combined with the Chebyshev polynomials basis, as described in **Table 1**. It featured the arrangement of a pair or triplet of atoms around the target atom.[67] The cut-off function employed in this work is the cosine function introduced by Behler and given in Eq. (1):[65]

| Non-local radial function | Non-local angular function |
|---|---|
| $G^2$ symmetry-function: $$G_i^2 = \sum_j e^{-\eta(r_{ij}-R_s)^2} \cdot f_{cut}(r_{ij})$$ Hyperparameters: $\eta, R_s$. | $G^3$ symmetry-function: $$G_i^3 = 2^{1-\xi} \sum_{j \neq i} \sum_{k \neq i,j} (1 + \lambda \cdot \cos(\theta_{jik}))^\xi \cdot f_{cut}(r_{ij}) \cdot f_{cut}(r_{ik})$$ Hyperparameters: $\xi, \lambda$. |
| First-order Chebyshev polynomials: $$T_0(x) = 1$$ $$T_1(x) = x$$ $$T_{l+1}(x) = 2x \cdot T_l(x) - T_{l-1}(x)$$ | Second-order Chebyshev polynomials: $$U_0(x) = 1$$ $$U_1(x) = 2x$$ $$U_l(x) = 2x \cdot T_l(x) - U_{l-1}(x)$$ From second-order polynomials, we build first-order pseudo-polynomials, $T_{l+\frac{1}{2}}$: $$\Lambda = \left(\frac{1+x}{2}\right)^{0.5}, \quad M = (1-x^2)^{0.5}, \quad N = \left(\frac{1-x}{2}\right)^{0.5}$$ $$T_{l+\frac{1}{2}}(x) = \Lambda \cdot T_l(x) - MN \cdot U_{l-1}(x)$$ |
| Perturbed $G^2$ symmetry-function: $$\tilde{G}_i^2 = \sum_{j \neq i} \sum_{k \neq i,j} T_l\left(\frac{r_{jk}}{R_c}\right) \cdot e^{-\eta(r_{jk})^2} \cdot f_{cut}(r_{jk})$$ Hyperparameters: $\eta, l, R_c$. Non-local information | Perturbed $G^3$ symmetry-function: $$\tilde{G}_i^3 = 2^{1-\xi} \sum_{j \neq i} \sum_{k \neq i,j} T_{l+\frac{1}{2}}(\cos(\theta_{ijk})) \cdot (1 + \lambda \cos(\theta_{jik}))^\xi \cdot f_{cut}(r_{ij}) \cdot f_{cut}(r_{ik})$$ Hyperparameters: $\xi, \lambda, l, R_c$. Non-local information |

**Table 1**: *Construction of the perturbed symmetry functions used to fingerprint non-local information. The hyperparameter η controls the width of the Gaussian function described by the $G^2$ function, and $R_s$ centres the Gaussian at the specified distance away from the atom target of the fingerprint. In the $G^3$ function, λ determines whether the cosine function is centered on 0 or π, and ξ controls the width of the function. In the non-local functions, $R_c$ is the cut-off radius of the fingerprint and l determines the order of the Chebyshev polynomial employed.*

$$f_{cut}(r_{ij}) = \begin{cases} 0.5 \cdot \left(\cos\left(\frac{r_{ij}}{R_c} \cdot \pi\right) + 1\right), & if\ r_{ij} \leq R_c \\ 0, & if\ r_{ij} > R_c \end{cases} \quad (1)$$

The prediction of atomic forces was decomposed into two parts: amplitude and direction. The force amplitude, which is translation and rotation invariant, and therefore, the symmetry and perturbed symmetry functions can be used for their predictions. The directional fingerprint, $G^D$, was used to determine the direction of the force. It is based on the $G^2$ symmetry function and described in Eq. (2):[42]

$$\boldsymbol{G}_i^D = \frac{1}{N_j} \sum_{j \neq i} \boldsymbol{r}_{ij} \cdot e^{-\eta(|r_{ij}|-R_s)^2} \cdot f_{cut}(|\boldsymbol{r}_{ij}|) \qquad (2)$$

Where $|\boldsymbol{r}_{ij}|$ represents the Euclidian norm of the vector $\boldsymbol{r}_{ij}$ from the fingerprinted atom $i$ to its neighbor $j$. $N_j$ represents the number of neighbors accounted for, and the factor $1/N_j$ is used to normalize the fingerprint.

To complete the fingerprint, information relevant to the chemical nature of the environment around each atom was captured through a collision-free weighting approach introduced by Beevers *et al.*[68] According to number theory, any given natural integer, $Z \in \mathbb{Z}$, can be expressed as a product of prime factors raised by an appropriate exponent in a unique fashion, as shown in Eq. (3):

$$Z = 1^0 \cdot 2^{a_1} \cdot 3^{a_2} \cdot 5^{a_2} \cdot \ldots \cdot p_n{}^{a_n} = \prod_{n=1}^{\infty} p_n{}^{a_n} \qquad (3)$$

Where $p_n$ are successive prime numbers and $a_n$ are the appropriate exponents. Passing Eq. (3) to the logarithmic, prime numbers form the basis of a vector space where the exponents are its coefficients, as shown in Eq. (4):

$$\log(Z) = \sum_{n=1}^{\infty} a_n \cdot x_n, \qquad x_n = \log(p_n) \qquad (4)$$

The chemical nature of elements in the fingerprint was described by associating them with a unique prime number, forming a basis set of the form $\boldsymbol{X} = \{p_1, p_2, \ldots, p_K\}$ for $K$-elements in the structure. Thus, the fingerprint reflected both the atom's nature and the nature of its closest neighbors. The weight associated with each atom is described by Eq. (5):

$$W = \omega + \widetilde{\omega} \qquad (5)$$

Where $\omega$ is the on-site weight, $\omega = p_i$, where $p_i$ is the prime number associated with the atom's element in the basis set $\boldsymbol{X}$ and $\widetilde{\omega}$ is the neighbor's weight contribution calculated using Eq. (4). The coefficients are the number of atoms directly coordinated to the target atom of a given element, multiplied by the element's basis vector in $\boldsymbol{X}$. The fingerprinting procedure is described in **Figure 1** and further details on the fingerprint employed for the three datasets investigated are described in detail in the SI.

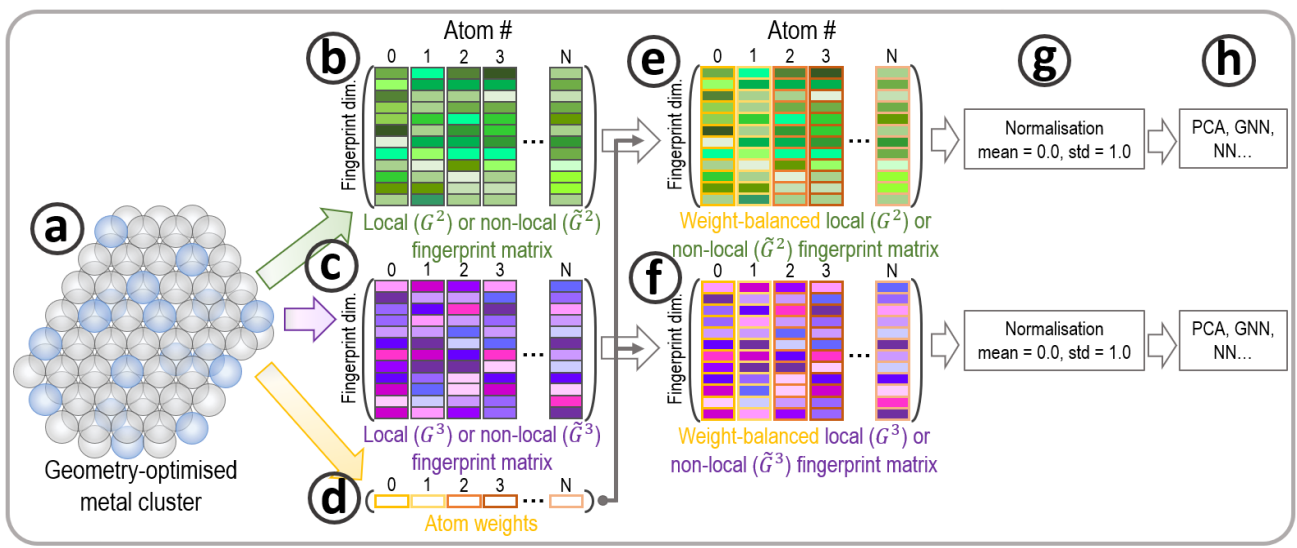

*Figure 1: Workflow of extracting the fingerprint of each atom in the system and applying atomic weight. From (a) DFT-optimized particles, the symmetry functions are employed to extract fingerprint matrices (b, c) and weight vectors for each atom (d). Each atomic weight is then applied to the fingerprint matrices (e, f), and those matrices are normalized (g) and fed to ML algorithms (h).*

### 2.4. Building the neural networks

#### 2.4.1. Graphical neural network: Energy prediction

The data used in this work covered a wide variety of cluster shapes and sizes, leading to different fingerprint arrays, i.e., non-tabular data, which graph neural networks (GNN) are known to work conveniently with data of this nature. The developed GNN required three elements to predict atomic energies: The fingerprint matrix, **M$_e$**, describing the atom's environments and forming the nodes of the graph; the adjacency matrix, **A**, describing the node connections within the graph; and the atom-weight matrix, **W**, capturing the element of each atom. Error! Reference source not found. reports the structures and parameters employed for the different GNNs built in this work, and we refer the reader to the PyTorch and PyTorch-Geometric libraries documentation for details on the different parameters used to build the different networks.[69,70]

| GNN structure | Optimizer | Weight and biases |
|---|---|---|
| Pd-pure:150-150-150-150-150-1,[a] AuPd-alloy: 400-150-150-150-150-1, Pd/silica: 150-150-150-150-150-150-1, no message passing. Global-add-pool readout function. Activation functions: RReLU (Pd-pure), Leaky-ReLU (AuPd-alloy), Leaky-RReLU (Pd/silica). | Pd-pure: NAdam, 80% α-dropout regularization, 660 epochs. AuPd-alloy: NAdam, L2-regularization strength: 0.5×10$^{-3}$, 960 epochs Pd/silica: NAdam, L2-regularization strength: 10$^{-3}$, 1200 epochs | Pd-pure, AuPd-alloy: Weights: Xavier uniform, Bias: $\mathcal{N}(\mu=0.0, \sigma=1.0)$. Pd/Silica: Weights: Kaiming normal, Bias: $\mathcal{N}(\mu=0.0, \sigma=1.0)$. |

***Table 2****: Detailed structure of the GNN built for energy predictions. a: Indicates the number of neurons in each layer i.e. a first, second, third, fourth, and fifth layer containing 150 neurons, and an output layer of 1 neuron.*

The predicted total energy of a cluster was calculated as a sum of individual atomic contributions independently of the cluster's shape, size, nature, or state (gas-phase or supported). During the learning process, all datasets were split 80% for training and 20% for validation. From the training set, 16% of the total dataset was used for on-the-fly validation, avoiding overfitting. The NNIPs defined in the present work were built using the PyTorch and PyTorch-geometric libraries,[69,70] and according to the workflow depicted in **Figure 2** and described as follows:

a) The chemical space is sampled using DFT. Datasets are created, and the fingerprints are constructed;

b) A graph neural network (GNN) is employed to predict the system's energy. Each node in the graph corresponds to a cluster atom with an associated energy. A readout function sums the energy predicted on each node according to Eq. (6), where $\varepsilon_n$ is the value attached to one node (atom) in the graph:

$$E_{GNN} = \sum_{n,nodes} \varepsilon_n \quad (6)$$

c) The combination of two feedforward NNs provides the atomic forces at every optimization step until convergence is reached.

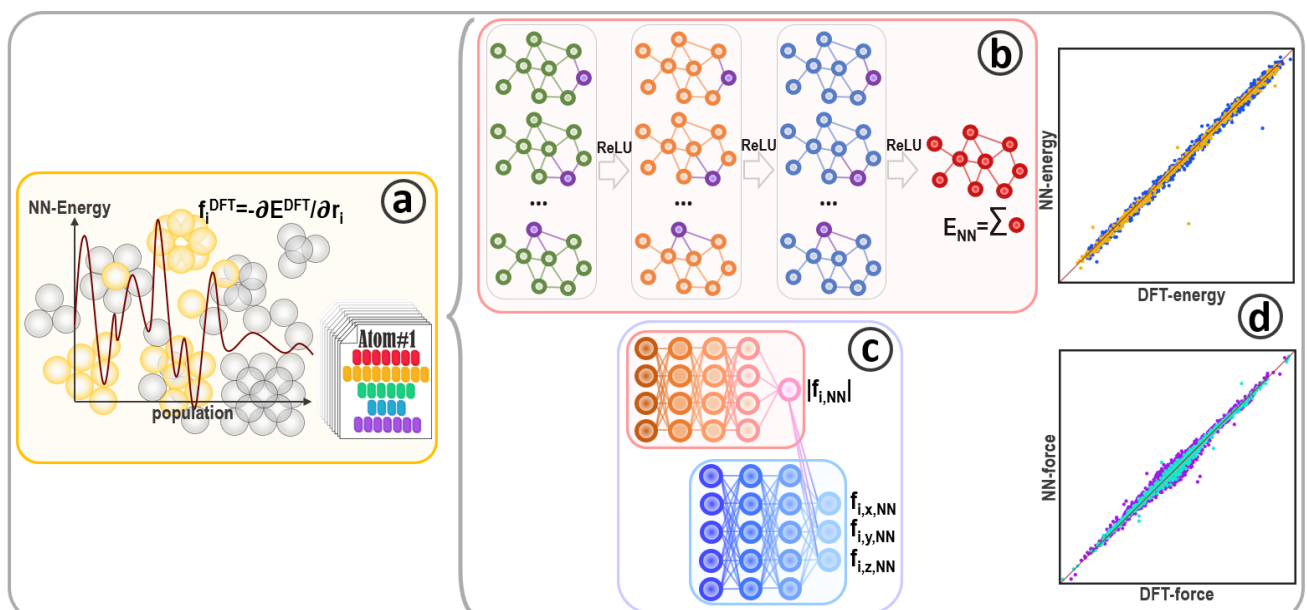

*Figure 2*: Schematic workflow of the predictive ML-calculator introduced in this work. (a) The dataset is generated. The dataset stores DFT total energies and atomic forces as targets, and the fingerprint of each atom is extracted. (b) Each fingerprint is assembled into a graph. Each node in the graph represents an atom, and a connection between two nodes represents a chemical bond. The GNN predicts atomic and total energies. (c) The second set of NNs predicts the Euclidian norm of the force acting on each atom and its direction. (d) The predictions are compared against the dataset data.

### 2.4.2. Feedforward neural network: Forces prediction

Classic feedforward networks were built for the prediction of the forces' norms and directions, as illustrated in **Figure 2**. The NNs' structures are described in **Table 3** and **Table 4**.

| Feedforward NN structure | Optimizer | Weight and biases |
|---|---|---|
| Pd-pure and AuPd-alloy:280-80-60-1, Pd/silica: 80-40-40-1, Activation functions: RReLU (Pd-pure and AuPd-alloy), Leaky-ReLU (Pd/silica). The softplus activation function was applied on the output layers of each NN. | All datasets: NAdam, L2-regularization strength: $0.5 \times 10^{-3}$, 1200 epochs. | All datasets: Weights: Xavier normal, Bias: $\mathcal{N}(\mu = 0.0, \sigma = 1.0)$. |

*Table 3*: Detailed structure of the feedforward NNs built for forces-norm predictions.

| Feedforward NN structure | Optimizer | Weight and biases |
|---|---|---|
| All datasets:400-120-100-100-100-3, Activation functions: Leaky-ReLU (all datasets). | All datasets: NAdam, L2-regularization strength: $10^{-4}$, 1000 epochs. | All datasets: Weights: Xavier uniform, Bias: Zeros. |

*Table 4*: Detailed structure of the GNN built for energy predictions.

## 3. Results

### 3.1. Data pre-processing

#### 3.1.1. Near-equilibrium structures and filters

The quality of the data employed during training is intrinsically linked to the quality of the prediction, and therefore, great care was taken regarding the dataset's pre-processing. In particular, the DFT cluster optimization led to a major proportion of the dataset describing quasi-identical near-equilibrium structures. In order to avoid overweighting quasi-identical images in the datasets, a filter that operates on each trajectory step was designed. The filter recursively compares the energy per atom of consecutive images: If the energy difference falls below a threshold, the current image is ignored and not included in the dataset.

For the gas-phase datasets, a threshold of $3\times10^{-3}$ eV/atom was chosen to capture the diversity in terms of fingerprint. As illustrated in **Figure 3** for the Pd gas-phase dataset, applying such a filter significantly reduces the number of points with norm forces between 0.00 and 0.40 eV/Å. Still, it does not influence those belonging to off-equilibrium structures. On supported Pd clusters, due to the complexity of this chemical space and rigidity of the support, a lower threshold of $1\times10^{-3}$ eV/atom was set to include a higher number of atomic environments close to the equilibrium. For consistency, the Pd-pure gas-phase clusters included in the Pd-$SiO_2$(001) dataset also had the same threshold.

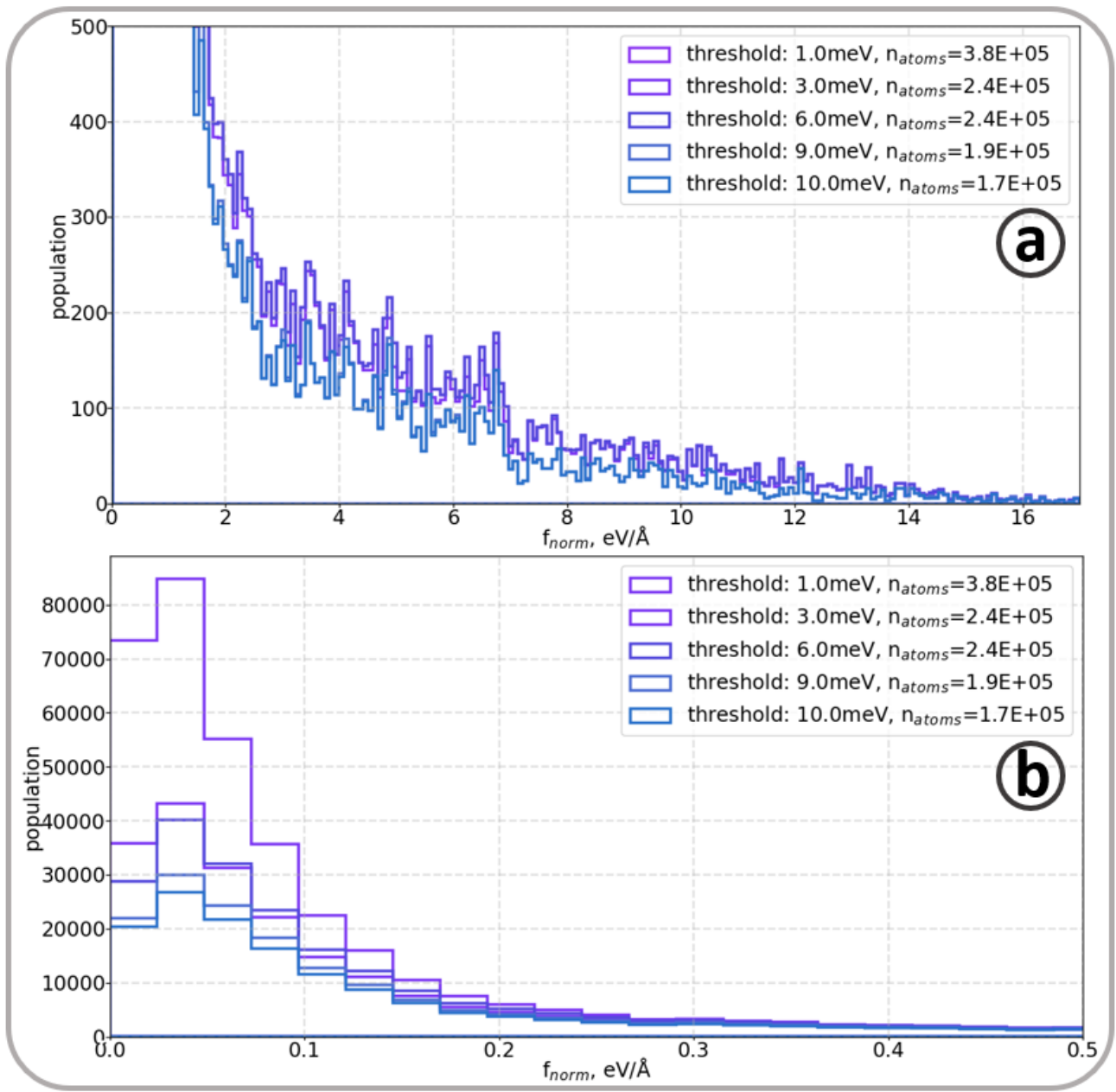

*Figure 3*: (a) Comparison of different filter thresholds (1, 3, 6, and 9 meV) effect on the number of atoms in the Pd-pure dataset for force predicting at small atomic force value (0.00-0.50 eV/Å). (b) Zoom-in on the near-zero forces region, i.e. the range [0.0, 0.5] eV/ Å.

### 3.1.2. Fingerprint noise reduction

In the scope of this work, the fingerprinting procedure can be considered as an information channel converting information from ase.Atoms object into a tensor object. According to information theory, this procedure introduces noise in the data, that can perturb the learning procedure and hinder the performance of ML-based algorithm.[71] To reduce the influence of the noise on the noise on the data's quality, different pre-processing methods were tested to improve the NN's predictions. Two commonly employed methods were compared, principal component analysis (PCA) and auto-encoders (AE) which proved powerful pre-processing tools.[72,73] PCA builds linear relationships between components by projecting a matrix of dimension $[N \times D]$ to a new matrix $[N \times D']$, where $N$ represents the number of atoms in a cluster, $D$ the initial dimension of the fingerprint, and each dimension $D'$ represents a linear component derived from the initial $D$-dimensional fingerprint with $D' < D$. AE follows the same principle as PCA but through a structure similar to neural networks, building more complex relationships than linear components as PCA does. Whereas PCA remains deterministic, AE must be trained to reduce efficiently the initial fingerprint's dimension, leading to the loss of noise existing in the original data. The most significant influence of pre-processing was observed for the prediction of forces norms and the results obtained using PCA and AE were compared only on Pd/SiO$_2$ and using a smaller 60D fingerprint using local and non-local descriptors in **Table 5** and **Table 6**.

|  | Weight | R$_s$ (Å) | η (Å²) | λ (∅) | ξ (∅) |
|---|---|---|---|---|---|
| $G^2$ | [2, 3, 5, 7] | [3.0, 3.5, 4.0, 4.5, 5.0, 5.5, 6.0] | [1.0, 3.0, 6.0] | ∅ | ∅ |
| $G^3$ | [2, 3, 5, 7] | ∅ | ∅ | [-1, +1] | [2.0, 8.0, 16.0] |

***Table 5***: *Parameters employed in the local symmetry functions to build the Pd/SiO$_2$ forces norm fingerprint.*

|  | Weight | η (Å²) | Cheby. Deg. | λ | ξ | Pseudo. Cheby. Deg. |
|---|---|---|---|---|---|---|
| $\widetilde{G}^2$ | [2, 3, 5, 7] | [1.0, 3.0, 6.0] | [2, 4, 6] | ∅ | ∅ | ∅ |
| $\widetilde{G}^3$ | [2, 3, 5, 7] | ∅ | ∅ | [-1, +1] | [1.0, 8.0, 16.0] | [3, 4, 5] |

***Table 6****: Parameters employed in the non-local perturbed functions to build the Pd/SiO$_2$ forces norm fingerprint.*

The results obtained after the reduction of the initial 60D fingerprint indicate better performance from PCA by 0.02 to 0.03 eV/Å on average against AE, , illustrated in **Figure 4**, and an improvement of 0.05 eV/ Å against the baseline result obtained without pre-processing of the initial 60D fingerprint. The AE preprocessing performs the encoding-decoding procedure with minimal error, as shown by the AE learning curve in **Figure 4(a)** where a RMSE below 0.01 is reached after just 200 epochs, indicating that the encoding-decoding procedure is performed with minimal loss of information through data-compression and the dimension-reduced vector contains all the relevant information stored in the initial, higher-dimension vector. However, despite this result, PCA produced better results and was employed in this work to pre-process the fingerprint associated with forces norm.

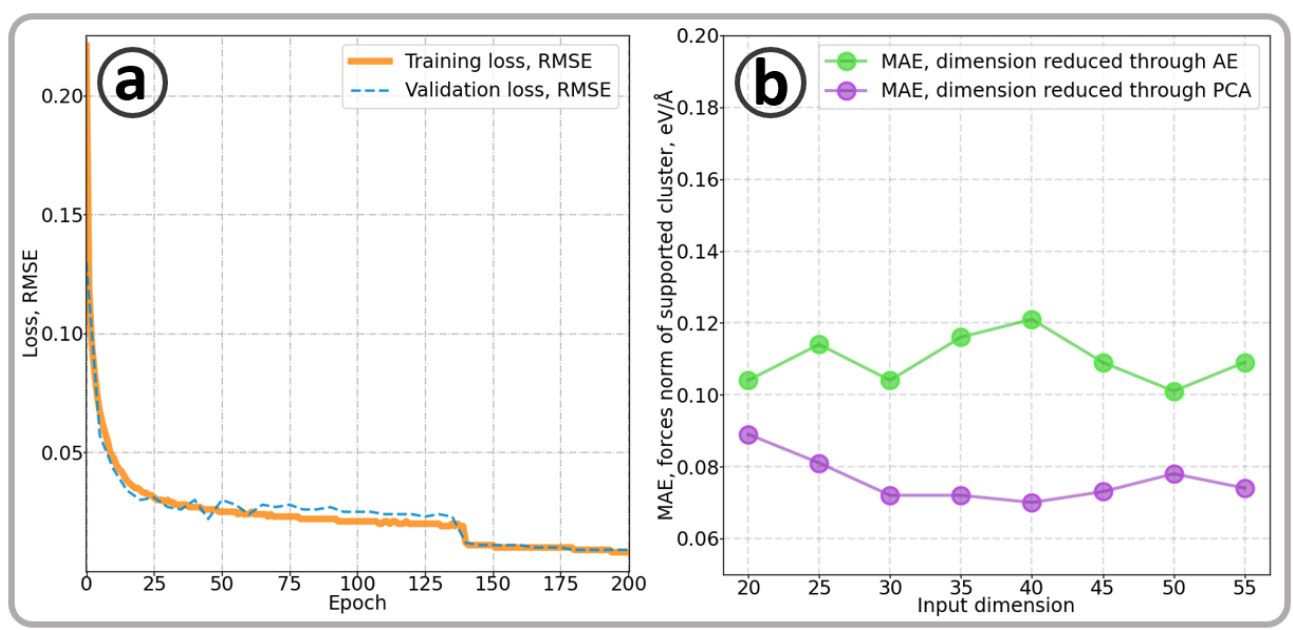

Figure 4: (a) The learning curve of the AE to reduce the dimension of the forces norm fingerprint from 60D to 30D. (b) The mean average error (MAE) of the neural network predicting the atomic forces norm with different pre-processing through auto-encoding (AE) and principal component analysis (PCA).

### 3.2. ML-calculator: Energy predictions

**Figure 5** shows the energy predictions and the mean absolute errors (MAE) for the three different systems: gas-phase Pd and AuPd nanoparticles and Pd/$\alpha$-SiO$_2$(001). The MAE ranges between 0.003 – 0.007 eV/atom and follows the order AuPd-alloy (gas) < Pd-pure (gas) < Pd/SiO$_2$. The most significant errors in the predicted energies correspond to unstable structures whose geometries are far from any minima in the potential energy surface. **[update RMSE]**

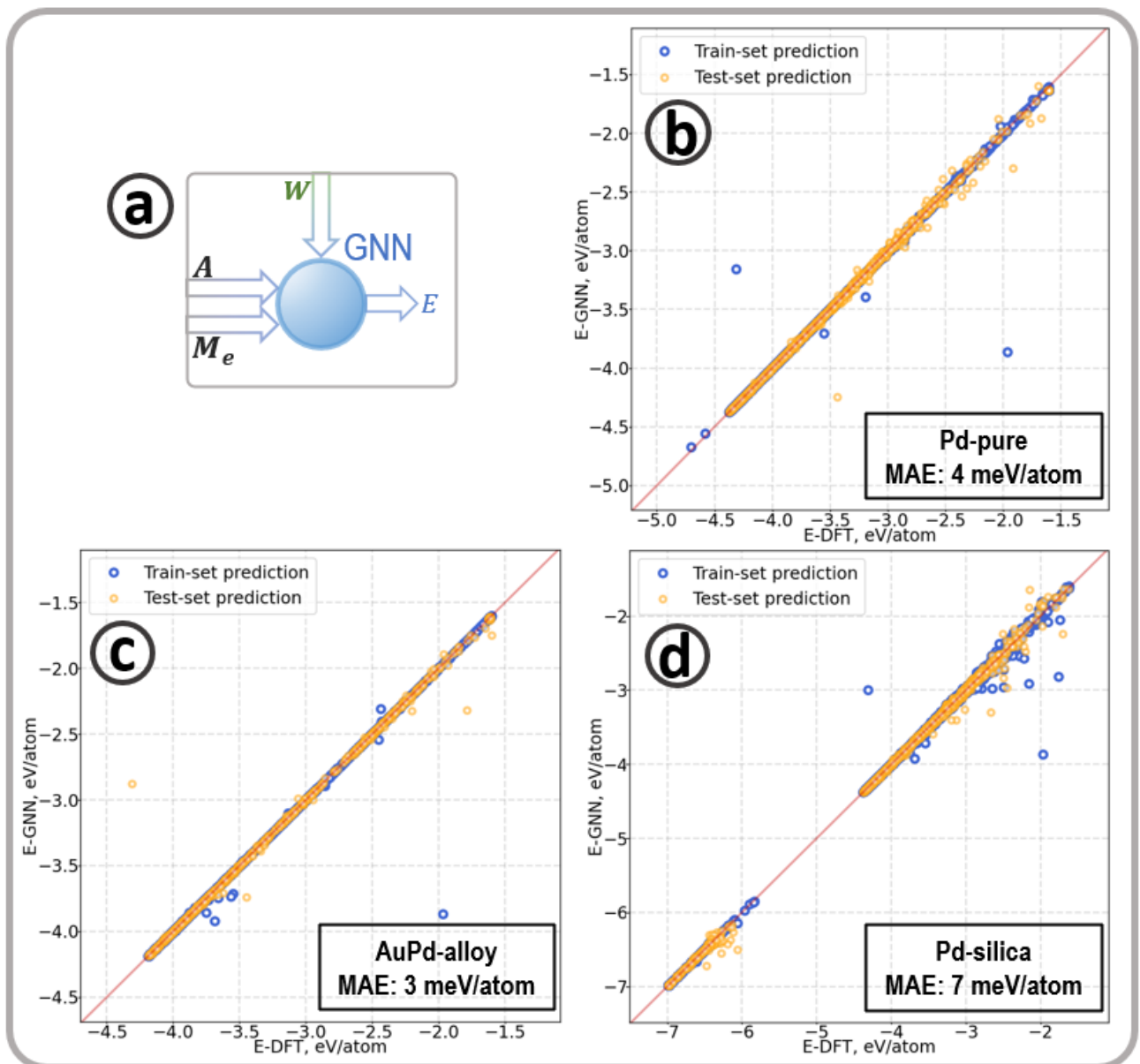

*Figure 5*: (a) Illustration of the part of the ML-calculator predicting the energy. The GNN requires the adjacency matrix (**A**), the energy fingerprint (**M**$_e$), and atom weights (**W**). (b), (c), and (d) show the correlation between predicted and DFT-energies for Pd-pure, AuPd-alloy, and Pd-silica datasets, respectively.

It is worth mentioning the flexibility of the ML-calculator to predict accurately the energy of alloy particles from a relatively small dataset containing actual multimetallic structures. **Figure 6** compares the predicted and DFT-calculated total energies of 12 particles containing 20 atoms with compositions in the range $Au_6Pd_{14}$ to $Au_{18}Pd_2$ with different configurations: Random, Janus, and core-shell. The GNN predicts total errors in the range of 0.001 to 0.017 eV/cluster (0.001 eV/atom).

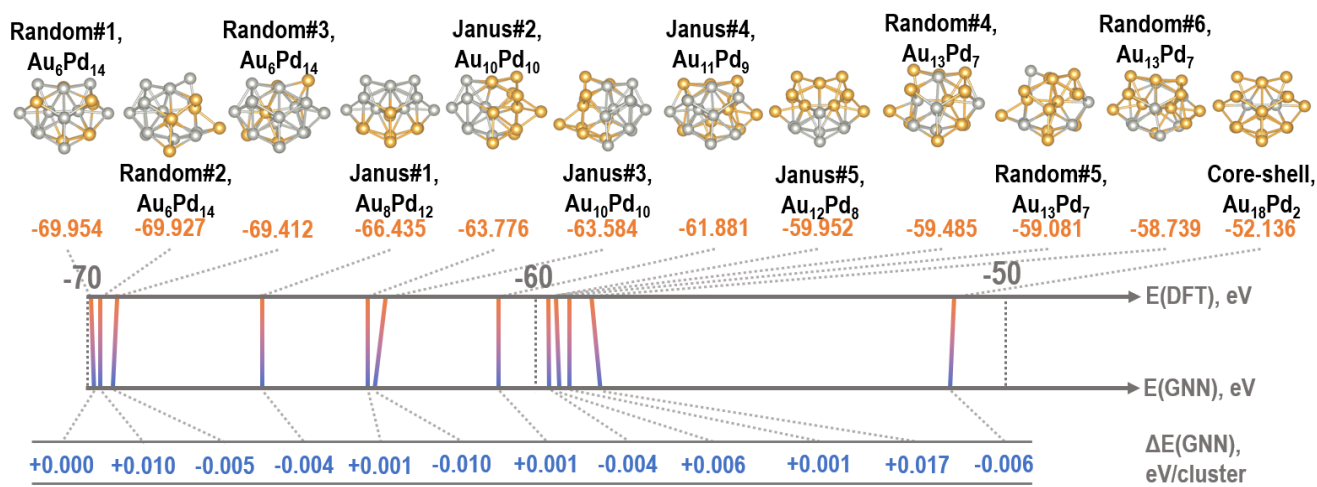

*Figure 6*: Diagrammatic representation of the bimetallic clusters containing 20 atoms. The DFT and GNN predicted total energies, as well as their difference, ΔE(GNN), are in eV over the entire tested cluster, i.e. eV/cluster.

### 3.3. ML-calculator: Forces predictions

Atomic force is the product of the force norm and the direction vector. **Figure 7(a)** describes the PCA-transformed forces-norm generated from a first NN feeding a second feedforward NN for the prediction of the vectorial forces. The accuracies of the predictions are summarized in **Table 7**, where $\delta E$ represents the energy threshold to select two images from the same optimization path (near-equilibrium pre-processing), MAE is the mean absolute error, and RMSE is the root mean squared error. Predictions of the vectorial forces consistently show an improvement compared to predictions of the forces norm, as illustrated in **Table 7** by reducing the MAE by 37%, 22%, and 44% for the Pd-pure, AuPd-alloy, and Pd/silica datasets, respectively. This observation can be explained mathematically: The norm is multiplied with a 3D vector whose component are inferior or equal to 1 (direction unit vector), generating smaller values that compared against DFT forces result in a smaller difference. Furthermore, because the direction vector is trained independently from the norm, the second NN can undirectly learn a correction to the force norm in each of the 3 vectorial coordinates and apply to the norm before the loss is calculated.

In the chosen structure of two embedded feedforward NNs, the direction (rotation-covariant) NN acts as the unit-vector prediction and corrects the forces-norm. It is also noticeable in **Table 7** that, the more accurate the 'guess' on forces norm, the higher the accuracy on the resultant vectorial force. The graphical comparison between predicted forces and DFT forces is in **Figure 7** (b-d) for the three systems understudy.

| Dataset | $\delta E$, meV/atom | Number of entries | Norm error, MAE, eV/Å | Vectorial force error, MAE, eV/Å | Vectorial force error, RMSE, eV/Å |
|---|---|---|---|---|---|
| Gas-phase Pd-pure | 3.0 | 240,000 | 0.080 | 0.050 | 0.112 |
| Gas-phase AuPd-alloy | 3.0 | 150,000 | 0.055 | 0.043 | 0.109 |
| Gas-phase and supported Pd/SiO$_2$ | 1.0 | 500,000 | 0.086 | 0.048 | 0.130 |

*Table 7: Predictions accuracy in forces norm and total vectorial forces for the three datasets understudied in this work. $\delta E$ is the energy threshold used in the pre-processing filter.*

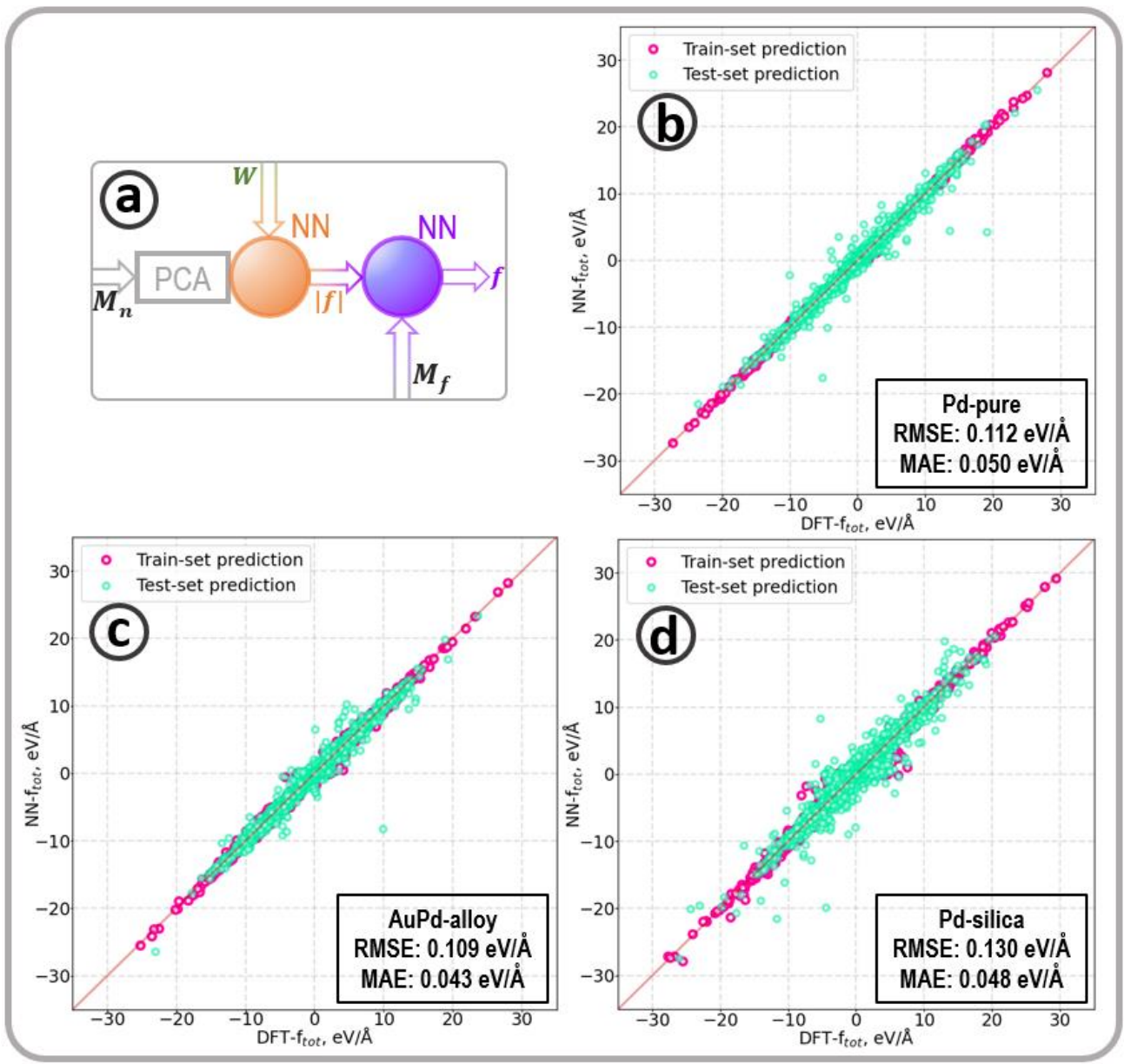

*Figure 7*: (a) Illustration of the part of the ML-calculator predicting atomic forces. A first feedforward NN predicts the forces' norm using the atomic weights multiplied by the PCA-reduced fingerprint, $M_n$. A second feedforward NN predicts the total force using the direction fingerprint for input, $M_f$, and multiplying the 3D output with the predicted forces norm. (b), (c), and (d) show the correlations between DFT and predicted forces for the Pd-pure, AuPd-alloy, and Pd-silica-dataset.

### 3.4. Interpretability of the results

The GNN is trained to provide a set of atomic energies with the only restriction that the sum of energies should approach the DFT energy of the system under study. This particularity raises the question of the interpretability of each atomic contribution in the final sum, i.e. does the GNN learn the 'physics' associated with the dataset, or is the predicted atomic energy a mathematical artifact with no physical value?

Four Pd-pure gas-phase clusters, illustrated in **Figure 8**, were chosen to predict the cohesion energy of specific atoms in the vertex, edges, and facets. The energy variation of pulling individual atoms away from the metal cluster can be approached with a Morse potential curve;[74,75] the general equation is in Eq. (8).

$$V(r) = \varepsilon\left[e^{-2a(r-r_e)} - 2e^{-a(r-r_e)}\right] \qquad (8)$$

Where $r_e$ represents the equilibrium distance between the target atom and the cluster, $\varepsilon$ is the potential that reached the bottom of the Morse curve well and quantifies the strength of the interaction between the atom and the rest of the cluster, i.e., the cohesion energy, $E_{coh}$. The parameter $a$ expresses the width of the well and is related to the stiffness of the interaction, $k_e$, at the bottom of the well, $a = \sqrt{k_e/2\varepsilon}$. In addition, the norm of the force vector predicted by the NN should equal the derivative of the Morse curve, expressed in Eq. (9).

$$\frac{dV}{dr} = -2\varepsilon a e^{a(r_{eq}-r)}\left(e^{a(r_{eq}-r)} - 1\right) \qquad (9)$$

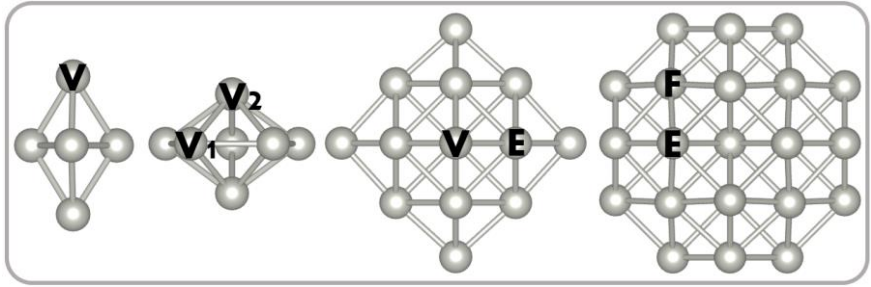

*Figure 8: Representation of the four Pd-pure clusters selected to evaluate the interpretability of the energy predictions. From left to right: $Pd_5$, $Pd_7$, $Pd_{19}$, and $Pd_{38}$. The initials V, E, and F indicate the position of the atoms selected to investigate their cohesion energy on the vertex, edge, and facet.*

The distances between the targeted atoms and the clusters were systematically increased, and the energies and forces were predicted. **Figure 9** (a) represents the predicted energies following the Morse curves, i.e. with coefficients of determination ($R^2$) at least 0.9. The meaning of correlation between predicted energies and well-accepted potentials indicates that the GNN learned the underlying physics carried by the atomic fingerprints. In other words, an interpretable physical meaning can be associated with each contribution to the total energy predicted by the GNN. Furthermore, the evaluation of the force norms predicted also shows good agreement ($R^2 \geq 0.93$) with the expected trend given by Eq. (9) and represented in **Figure 9(b)**, demonstrating the interpretability of the ML-calculator.

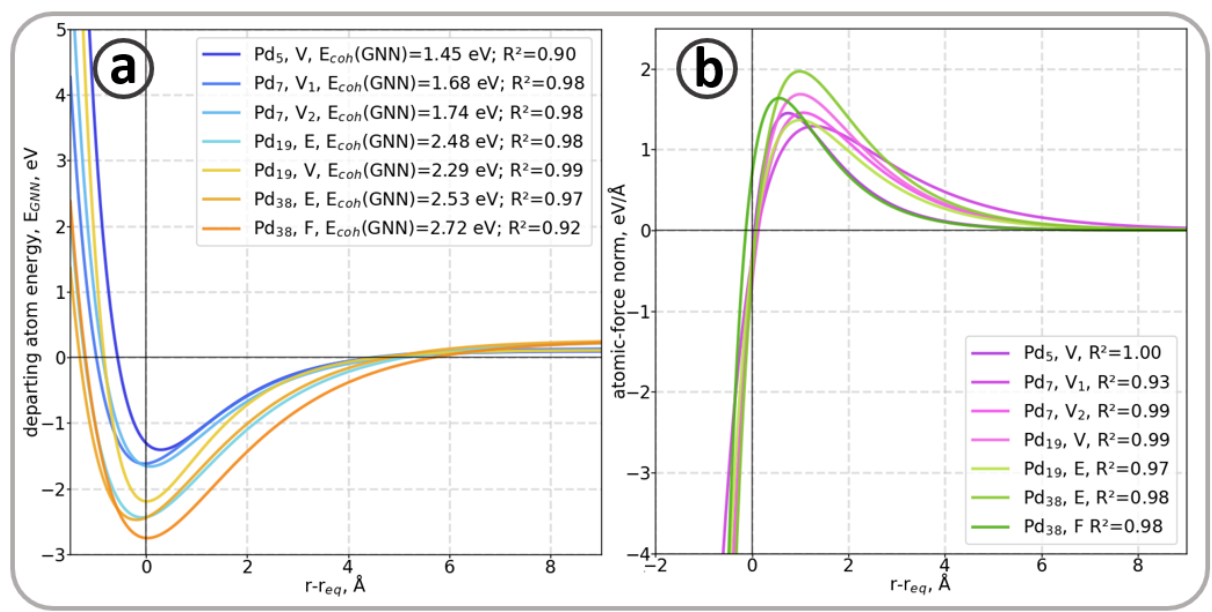

**Figure 9**: (a) Cohesion energy and (b) force's norm of the targeted atoms in $Pd_x$ clusters (x=5, 7, 19, 38) as a function of the atom-cluster distance.

## 4. Conclusion

The work presents an improved atomic cluster fingerprinting able to capture local, non-local, and the nature of atoms matter, easing the use of advanced computational techniques in physical science, particularly nanoscience and catalysis. The fingerprint feeds three different machine learning structures to accurately predict atomic and cluster energies and atomic force norms and directions. Following an energy-free approach, these graphical and feedforward networks were combined in an autonomous machine-learning calculator.[42] This calculator was tested against gas-phase Pd, AuPd, and $Pd/SiO_2$ clusters, representing contemporary challenges to designing multimetallic and supported catalysts. Analysis of the energy and forces predictions revealed a near DFT accuracy for the different systems. Besides, the atomic energy interpretability was tested and confirmed to encapsulate physical meaning, such as the cohesion energy. Overall, the innovative ML-calculator is accurate and highly flexible, producing competitive results or better than existing neural networks' interatomic potentials (NNIPs).[35,39,40,76–81]